\documentclass[12pt]{article}

\usepackage[letterpaper,margin=0.75in]{geometry}
\usepackage{setspace}
\usepackage{fontspec}
\usepackage{graphicx}
\usepackage{amsmath,bm,amsbsy,amssymb}
\usepackage[dvipsnames]{xcolor}
\usepackage[labelformat=simple]{subcaption}

\usepackage{algorithm}
\usepackage{algpseudocode}
\usepackage[noadjust]{cite}
\usepackage{comment}
\usepackage{soul}
\usepackage{tikz}
\usepackage{pgfplots}
\usetikzlibrary{arrows.meta}
\pgfplotsset{compat=1.18}

\newcommand{\universalbeamforming}{
\begin{tikzpicture}[
    >=Latex,
    line/.style={line width=1.2pt},
    mult/.style={
        draw=black,
        circle,
        minimum size=0.42cm,
        inner sep=0pt,
        font=\small
    },
    add/.style={
        draw=black,
        circle,
        minimum size=0.42cm,
        inner sep=0pt,
        font=\small
    },
    outbox/.style={
        draw=blue,
        very thick,
        rectangle,
        minimum width=1.65cm,
        minimum height=0.7cm,
        font=\large,
        text=blue
    },
    postbox/.style={
        draw=black,
        very thick,
        rectangle,
        minimum width=5.7cm,
        minimum height=1.05cm,
        align=center,
        font=\small
    }
]

\definecolor{branchone}{RGB}{150,0,150}
\definecolor{branchtwo}{RGB}{255,0,0}
\definecolor{branchk}{RGB}{255,130,0}

\node at (-6.05+0.75,0.1) {\Large $\vdots$};

\draw[very thick] (-6.05,2.3-0.4) rectangle (-6.05 + 1.5,2.3+0.4);

\draw[very thick] (-6.05,1.35-0.4) rectangle (-6.05+1.5,1.35+0.4);

\draw[very thick] (-6.05,-1.1-0.4) rectangle (-6.05+1.5,-1.1+0.4);

\node[align=center] at (-6.05+0.75,-1.1) {$\boldsymbol{\Phi}_K$};

\node[align=center] at (-6.05+0.75,1.35) {$\boldsymbol{\Phi}_2$};

\node[align=center] at (-6.05+0.75,2.3) {$\boldsymbol{\Phi}_1$};

\node[
    draw,
    align=center,
    font=\scriptsize,
    inner sep=3pt
] at (-6.05+0.75,-2.4)
{$\Theta = \{\theta_1,\theta_2,\ldots,\theta_K \}$};
\draw[->,dashed] (-6.05+0.75,-2.1) -- (-6.05+0.75,-1.5);

\node[left] at (-8.05,0.5) {$\mathbf{v}[n]$};

\node[right,font=\scriptsize] at (-6.05+1.5,2.5) {$\boldsymbol{\Phi}_1^H\mathbf{v}[n]$};
\node[right,font=\scriptsize] at (-6.05+1.5,1.55) {$\boldsymbol{\Phi}_2^H\mathbf{v}[n]$};
\node[right,font=\scriptsize] at (-6.05+1.5,-0.9) {$\boldsymbol{\Phi}_K^H\mathbf{v}[n]$};

\draw[line, ->] (-8.05,0.5) -- (-8.05+1,0.5);
\draw[line, -] (-8.05+1,-1.1) -- (-8.05+1,2.3);
\draw[line, ->] (-8.05+1,2.3) -- (-6.05,2.3);
\draw[line, ->] (-8.05+1,1.35) -- (-6.05,1.35);
\draw[line, ->] (-8.05+1,-1.1) -- (-6.05,-1.1);
\node[mult] (m1) at (-1.75,2.3) {$\times$};
\node[mult] (m2) at (-0.55,1.35) {$\times$};
\node[mult] (mK) at (0.65,-1.1) {$\times$};

\node (a1) at (2.15,2.3) {};
\node (a2) at (2.15,1.35) {};
\node (aK) at (2.15,-1.1) {};

\node (out) at (4.25,0.75)
{$\hat{x}[n]$};

\draw[line, branchone, ->] (-6.05+1.5,2.3) -- (m1.west);
\draw[line, branchtwo, ->] (-6.05+1.5,1.35) -- (m2.west);
\draw[line, branchk, ->] (-6.05+1.5,-1.1) -- (mK.west);

\draw[line, branchone, ->] (m1.east) -- (a1.west);
\draw[line, branchtwo, ->] (m2.east) -- (a2.west);
\draw[line, branchk, ->] (mK.east) -- (aK.west);

\draw[very thick] (2.,-1.5) rectangle (2+1,2.7);

\node[align=center] at (2.5,0.75) {\Large $\sum$};

\draw[line, black, ->] (3,0.75) -- (out.west);

\node[postbox] (post) at (-1.75,-3.75)
{
$
\displaystyle
\mu_i[n]
=
\frac{
P_i\exp\!\left(
-\frac{1}{2\nu}\mathcal{L}\left(\mathbf{v}_0^{n-1},\boldsymbol{\Phi}_i\right)
\right)
}{
\sum_{j=1}^{K}
P_j\exp\!\left(
-\frac{1}{2\nu}\mathcal{L}\left(\mathbf{v}_0^{n-1},\boldsymbol{\Phi}_j\right)
\right)
}
$
};

\node (obs) at (-6.65,-3.75) {$\mathbf{v}_0^{n-1}$};
\draw[line, black, ->] (obs.east) -- (post.west);

\draw[line, branchone, ->] (-1.75,-3.05) -- (-1.75,2.1);
\draw[line, branchtwo, ->] (-0.55,-3.05) -- (-0.55,1.15);
\draw[line, branchk, ->] (0.65,-3.05) -- (0.65,-1.35);

\node[branchone] at (-1.15,1.8) {$\mu_1[n]$};
\node[branchtwo] at (0.05,0.78) {$\mu_2[n]$};
\node[branchk] at (1.3,-1.65) {$\mu_K[n]$};

\end{tikzpicture}
}

\newcommand{\universalequalizer}{
\begin{tikzpicture}[
    >=Latex,
    line/.style={line width=1.2pt},
    mult/.style={
        draw=black,
        circle,
        minimum size=0.42cm,
        inner sep=0pt,
        font=\small
    },
    outbox/.style={
        draw=blue,
        very thick,
        rectangle,
        minimum width=1.65cm,
        minimum height=0.7cm,
        font=\large,
        text=blue
    },
    postbox/.style={
        draw=black,
        very thick,
        rectangle,
        minimum width=5.9cm,
        minimum height=1.05cm,
        align=center,
        font=\small
    }
]

\definecolor{branchone}{RGB}{150,0,150}
\definecolor{branchtwo}{RGB}{255,0,0}
\definecolor{branchk}{RGB}{255,130,0}

\node[left] at (-8.05-0.5,0.5) {$\mathbf{v}(nT_s)$};

\draw[line, ->] (-8.05-0.5,0.5) -- (-7.05-0.5,0.5);
\draw[line, -] (-7.05-0.5,-1.1) -- (-7.05-0.5,2.3);
\draw[line, ->] (-7.05-0.5,2.3) -- (-6.05-0.5,2.3);
\draw[line, ->] (-7.05-0.5,1.35) -- (-6.05-0.5,1.35);
\draw[line, ->] (-7.05-0.5,-1.1) -- (-6.05-0.5,-1.1);

\draw[very thick] (-6.05-0.5,1.9) rectangle (-4.55-0.5,2.7);
\draw[very thick] (-6.05-0.5,0.95) rectangle (-4.55-0.5,1.75);
\draw[very thick] (-6.05-0.5,-1.5) rectangle (-4.55-0.5,-0.7);

\node[align=center] at (-5.30-0.5,2.3) {$\boldsymbol{\Psi}_1$};
\node[align=center] at (-5.30-0.5,1.35) {$\boldsymbol{\Psi}_2$};
\node[align=center] at (-5.30-0.5,-1.1) {$\boldsymbol{\Psi}_K$};

\node[
    draw,
    align=center,
    font=\scriptsize,
    inner sep=3pt
] at (-5.30-0.5,-2.4)
{$\Theta = \{\theta_1,\theta_2,\ldots,\theta_K \}$};
\draw[->,dashed] (-5.30-0.5,-2.1) -- (-5.30-0.5,-1.5);

\node[right,font=\scriptsize] at (-5.30+0.2,2.6) {$\tilde{y}_1(nT_s)$};
\node[right,font=\scriptsize] at (-5.30+0.2,1.65) {$\tilde{y}_2(nT_s)$};
\node[right,font=\scriptsize] at (-5.30+0.2,-0.8) {$\tilde{y}_K(nT_s)$};

\node at (-5.30-0.5,0.1) {\Large $\vdots$};

\draw[very thick] (-3.75,1.9) rectangle (-2.25,2.7);
\draw[very thick] (-3.75,0.95) rectangle (-2.25,1.75);
\draw[very thick] (-3.75,-1.5) rectangle (-2.25,-0.7);

\node[align=center] at (-3.00,2.3) {DFE $1$};
\node[align=center] at (-3.00,1.35) {DFE $2$};
\node[align=center] at (-3.00,-1.1) {DFE $K$};

\node at (-3.00,0.1) {\Large $\vdots$};

\draw[line, branchone, ->] (-4.55-0.5,2.3) -- (-3.75,2.3);
\draw[line, branchtwo, ->] (-4.55-0.5,1.35) -- (-3.75,1.35);
\draw[line, branchk, ->] (-4.55-0.5,-1.1) -- (-3.75,-1.1);

\node[mult] (m1) at (-0.55,2.3) {$\times$};
\node[mult] (m2) at (0.45,1.35) {$\times$};
\node[mult] (mK) at (1.45,-1.1) {$\times$};

\node[right,font=\scriptsize] at (-2.25,2.5) {$\hat{d}_1[n]$};
\node[right,font=\scriptsize] at (-2.25,1.55) {$\hat{d}_2[n]$};
\node[right,font=\scriptsize] at (-2.25,-0.9) {$\hat{d}_K[n]$};

\draw[line, branchone, ->] (-2.25,2.3) -- (m1.west);
\draw[line, branchtwo, ->] (-2.25,1.35) -- (m2.west);
\draw[line, branchk, ->] (-2.25,-1.1) -- (mK.west);

\node[left] (a1) at (3,2.3) {};
\node[left] (a2) at (3,1.35) {};
\node[left] (aK) at (3,-1.1) {};

\draw[line, branchone, ->] (m1.east) -- (a1.west);
\draw[line, branchtwo, ->] (m2.east) -- (a2.west);
\draw[line, branchk, ->] (mK.east) -- (aK.west);

\draw[very thick] (2.75,-1.5) rectangle (3.75,2.7);
\node[align=center] at (3.25,0.75) {\Large $\sum$};

\node (out) at (4.95,0.75) {$\hat{d}[n]$};
\draw[line, black, ->] (3.75,0.75) -- (out.west);

\node[postbox] (post) at (-0.75,-3.9)
{
$
\displaystyle
\mu_i[n]
=
\frac{
P_i\exp\!\left(
-\frac{1}{2\nu}\mathcal{L}\left(\mathbf{e}_{i,0}^{n-1}\right)
\right)
}{
\sum_{j=1}^{K}
P_j\exp\!\left(
-\frac{1}{2\nu}\mathcal{L}\left(\mathbf{e}_{j,0}^{n-1}\right)
\right)
}
$
};

\node (obs) at (-5.5,-3.9) {$\mathbf{e}_{i,0}^{n-1}$};
\draw[line, black, ->] (obs.east) -- (post.west);

\draw[line, branchone, ->] (-0.55,-3) -- (-0.55,2.05);
\draw[line, branchtwo, ->] (0.45,-3) -- (0.45,1.10);
\draw[line, branchk, ->] (1.45,-3) -- (1.45,-1.35);

\node[branchone] at (0.05,1.8) {$\mu_1[n]$};
\node[branchtwo] at (1,0.78) {$\mu_2[n]$};
\node[branchk] at (2.1,-1.65) {$\mu_K[n]$};

\end{tikzpicture}
}

\newcommand{\plotgeometryMACE}[1]{

\tikzset{every picture/.style={line width=0.75pt}}

\begin{tikzpicture}[x=0.75pt,y=0.75pt,yscale=-1,xscale=1]

\draw[color={rgb,255:red,255; green,255; blue,255},
      fill={rgb,255:red,255; green,255; blue,255}]
(199.5,455.23) -- (799.25,455.23) -- (799.25,774.23) -- (199.5,774.23) -- cycle;

\draw[color={rgb,255:red,74; green,144; blue,226},
      line width=1.5,
      line join=round,
      line cap=round]
plot[smooth, domain=204:773, samples=60]
(\x,{494 + 5*sin(4*\x) + 2.5*sin(9*\x)});

\draw[line width=1.5] (244,492.84) -- (243,691.84);

\draw[color={rgb,255:red,65; green,117; blue,5},
      fill={rgb,255:red,65; green,117; blue,5},
      draw opacity=1]
(226.98,493.3) -- (227,464.67)
.. controls (227,460.6) and (234.6,457.3) ..
(243.96,457.31)
.. controls (253.33,457.32) and (260.92,460.63) ..
(260.91,464.7)
-- (260.89,493.33) -- (243.93,500.69) -- cycle;

\draw (243.94,479) node[color=black,align=center, font=\small] {Array};

\foreach \y in {637.49,650.49,688.49}{
\draw[color={rgb,255:red,208; green,2; blue,27},
      fill={rgb,255:red,210; green,20; blue,20},
      fill opacity=0.98]
(242.74,\y) circle [x radius=5.02, y radius=5.02];
}

\draw[draw opacity=0,
      fill={rgb,255:red,245; green,166; blue,35},
      fill opacity=0.61]
(290.75,577.9)
.. controls (301.91,591.14) and (309.05,612.75) ..
(308.78,637.09)
.. controls (308.52,660.16) and (301.65,680.65) ..
(291.18,693.78) -- cycle;

\draw[draw opacity=0,
      fill={rgb,255:red,245; green,166; blue,35},
      fill opacity=0.61]
(242.77,637.49) -- (291.78,580.86) -- (292.18,693.78) -- cycle;

\draw[dash pattern={on 1.69pt off 2.76pt}, line width=1.5]
(242.77,637.49) -- (308.25,562.23);

\draw[dash pattern={on 1.69pt off 2.76pt}, line width=1.5]
(242.77,637.49) -- (305.25,708.23);

\draw[color={rgb,255:red,139; green,87; blue,42},
      draw opacity=0.7,
      line width=2.25]
(208.52,762.63) -- (773.1,762.63);

\draw[<->]
(454.11,608.38)
.. controls (457.31,612.79) and (459.23,618.45) ..
(459.23,624.61)
.. controls (459.23,629.51) and (458.02,634.09) ..
(455.91,637.98);

\draw (429.75,625.61) node[font=\Large] {$\theta_0$};

\draw (312.34,533.21) node [align=left] {Field-of-view};

\pgfmathsetmacro{\txshift}{0.35*#1}
\draw (491.47,591.5) node[rotate=-351.92,align=left] {direct path};
\draw (405.47,557.5) node[rotate=-336.99,align=left] {surface bounce path};
\draw (510.47,706.5) node[rotate=-325.75,align=left] {bottom bounce path};
\draw[<-,color={rgb,255:red,208; green,2; blue,27}, line width=1.5]
(261.5,633.86) -- ({729.25+\txshift},569.21);

\draw[<-,color={rgb,255:red,0; green,0; blue,0}, line width=1.5]
(265.95,626.71) -- (579.08,497.33) -- ({729.97+\txshift},569.03);

\draw[<-,color={rgb,255:red,0; green,0; blue,0}, line width=1.5]
(262.62,645.39) -- (445.38,762.63) -- ({729.97+\txshift},569.03);

\draw[dash pattern={on 0.84pt off 2.51pt}]
({729.97+\txshift},569.03) -- ({729.7+\txshift},642.44);
\draw[dash pattern={on 0.84pt off 2.51pt},<->]
(244.77,637.49) -- ({724.38+\txshift},637.49);

\draw (228.37,563.21) node[font=\large] {$D_R$};
\draw (708.77,523.21) node[font=\large] {$D_T$};
\draw (575.23,623.98) node[font=\large] {$L_{TR}$};

\begin{scope}[shift={({0.35*#1},0)}]

\draw[line width=1.5] (730.23,468.62) -- (729.97,549.03);

\draw[fill={rgb,255:red,74; green,144; blue,226}, fill opacity=1]
(720.4,554.93)
.. controls (720.4,551.45) and (723.22,548.63) ..
(726.7,548.63)
-- (734.7,548.63)
.. controls (738.18,548.63) and (741,551.45) ..
(741,554.93)
-- (741,561.23)
-- (720.4,561.23)
-- cycle;

\draw[fill={rgb,255:red,245; green,166; blue,35}, fill opacity=1]
(717.4,568.11)
.. controls (717.4,561) and (723.44,555.23) ..
(730.9,555.23)
.. controls (738.36,555.23) and (744.4,561) ..
(744.4,568.11)
.. controls (744.4,575.22) and (738.36,580.98) ..
(730.9,580.98)
.. controls (723.44,580.98) and (717.4,575.22) ..
(717.4,568.11) -- cycle;

\draw[fill={rgb,255:red,208; green,2; blue,27}, fill opacity=1]
(767.06,477.71) -- (743.97,496.98) -- (716.71,497.28) -- (693.19,478.52) -- cycle;

\draw[fill={rgb,255:red,74; green,144; blue,226}, fill opacity=1]
(713.53,461.86) -- (730.4,449.95) -- (730.66,472.58) -- cycle;

\draw (740.4,465.15) -- (770.15,465.15);
\node [align=center, font=\scriptsize] at (800.15,465.15) {direction of \\ motion};
\draw[shift={(770.15,465.15)}, rotate=180]
(10.93,-3.29)
.. controls (6.95,-1.4) and (3.31,-0.3) ..
(0,0)
.. controls (3.31,0.3) and (6.95,1.4) ..
(10.93,3.29);

\draw (729.97,569.03) node[align=left] {TX};

\end{scope}
\end{tikzpicture}
}

\usepackage{acronym}
\newacro{adc}[ADC]{analog-to-digital converter}
\newacro{adcs}[ADCs]{analog-to-digital converters}
\newacro{dac}[DAC]{digital-to-analog converter}
\newacro{dacs}[DACs]{digital-to-analog converters}
\newacro{snr}[SNR]{signal-to-noise ratio}
\newacro{ber}[BER]{bit error rate}
\newacro{ser}[SER]{symbol error rate}
\newacro{mse}[MSE]{mean squared error}
\newacro{avb}[AVB]{audio video bridging}
\newacro{ofdm}[OFDM]{orthogonal frequency division multiplexing}
\newacro{zpofdm}[ZP-OFDM]{zero-padding OFDM}
\newacro{cpofdm}[CP-OFDM]{cyclic prefix OFDM}
\newacro{dsofdm}[DS-OFDM]{direct sequence OFDM}
\newacro{psk}[PSK]{phase-shift keying}
\newacro{qpsk}[QPSK]{quaternary PSK}
\newacro{bpsk}[BPSK]{binary PSK}
\newacro{qam}[QAM]{quadrature amplitude modulation}
\newacro{fft}[FFT]{fast Fourier transform}
\newacro{idft}[IDFT]{inverse discrete Fourier transform}
\newacro{dft}[DFT]{discrete Fourier transforms}
\newacro{ls}[LS]{least squares}
\newacro{mrc}[MRC]{maximum ratio combining}
\newacro{dmrc}[DMRC]{differential maximum ratio combining}
\newacro{crc}[CRC]{cyclic redundancy check}
\newacro{spl}[SPL]{sound pressure level}
\newacro{psd}[PSD]{power spectral density}
\newacro{ue}[UE]{user equipment}
\newacro{bs}[BS]{base station}
\newacro{aoa}[AoA]{angles of arrival}
\newacro{act}[ACT]{acoustic communications testbed}
\newacro{sdma}[SDMA]{spatial division multiple access}
\newacro{ici}[ICI]{inter-carrier interference}
\newacro{foc}[FOC]{Frequency Offset Compensation}
\newacro{cdf}[CDF]{cumulative distribution function}
\newacro{dfe}[DFE]{decision-feedback equalizer}
\newacro{isi}[ISI]{inter-symbol interference}
\newacro{pll}[PLL]{phase-locked loop}
\newacro{dll}[DLL]{delay-locked loop}
\newacro{rls}[RLS]{recursive least squares}
\newacro{lms}[LMS]{least mean squares}
\newacro{lms}[LMS]{least mean squares}
\newacro{tr}[TR]{time-reversal}
\newacro{pc}[PC]{phase-conjugate}
\newacro{simo}[SIMO]{single-input multiple-output}
\newacro{mimo}[MIMO]{multiple-input multiple-output}
\newacro{csi}[CSI]{channel state information}
\newacro{mmse}[MMSE]{minimum mean-square error}

\title{\LARGE \bf Universal Adaptive Beamforming: A Bayesian Approach}

\author{\small
Diego A. Cuji\textsuperscript{1},
Andrew C. Singer\textsuperscript{1},
and John R. Buck\textsuperscript{2}\\[0.5em]
\textsuperscript{1}\small Stony Brook University, Stony Brook, NY, USA\\
\textsuperscript{2}\small University of Massachusetts Dartmouth, Dartmouth, MA, USA
}

\date{}

\begin{document}

\maketitle

\begin{abstract}
We present a Bayesian universal beamforming framework for adaptive array processing in dynamic underwater acoustic environments with unknown and time-varying propagation geometry. Motivated by ideas from universal prediction and estimation, the proposed approach discretizes the angular domain into a finite set of steering hypotheses and recursively computes posterior probabilities over competing spatial models using observation-dependent likelihood functions. For Gaussian observation models, the posterior update reduces to an exponential-weights recursion driven by hypothesis-dependent beamformer evidence metrics. The resulting framework performs soft spatial inference and adaptive beamforming by continuously redistributing posterior probability across competing steering hypotheses while forming posterior-weighted combinations of branch outputs. The formulation naturally connects to classical adaptive beamformers including matched filtering and minimum mean-square error (MMSE) beamforming.

In addition, the framework is extended toward broadband underwater acoustic communication receivers through frequency-domain beamformer synthesis and adaptive equalization. Posterior probabilities are updated according to branch-specific equalization errors, enabling joint spatial-temporal adaptation under multipath propagation, Doppler-induced distortions, and time-varying channel conditions. Experimental results using MACE data demonstrate reliable communication performance with low overhead, low data detection mean-squared error, and zero observed bit errors.


\end{abstract}

\section{Introduction}


Underwater acoustic communication channels exhibit severe multipath propagation, long delay spreads, Doppler effects, and strong temporal variability. Surface and bottom reflections generate arrivals with different delays, gains, and angles of arrival, while transmitter-receiver motion introduces Doppler distortion. Consequently, underwater acoustic communication systems must operate in environments where both the temporal and spatial characteristics of the propagation channel vary over time. 

\begin{figure}[h]
    \centering
    \begin{subfigure}[b]{0.8\textwidth}
    \centering
    \begin{subfigure}[b]{0.64\textwidth}
        \centering
        \scalebox{0.48}{\plotgeometryMACE{0}}
        \caption{}
        \label{fig:geometry}
    \end{subfigure}
    \begin{subfigure}[b]{0.34\textwidth}
        \centering
        \includegraphics[width=0.75\textwidth]{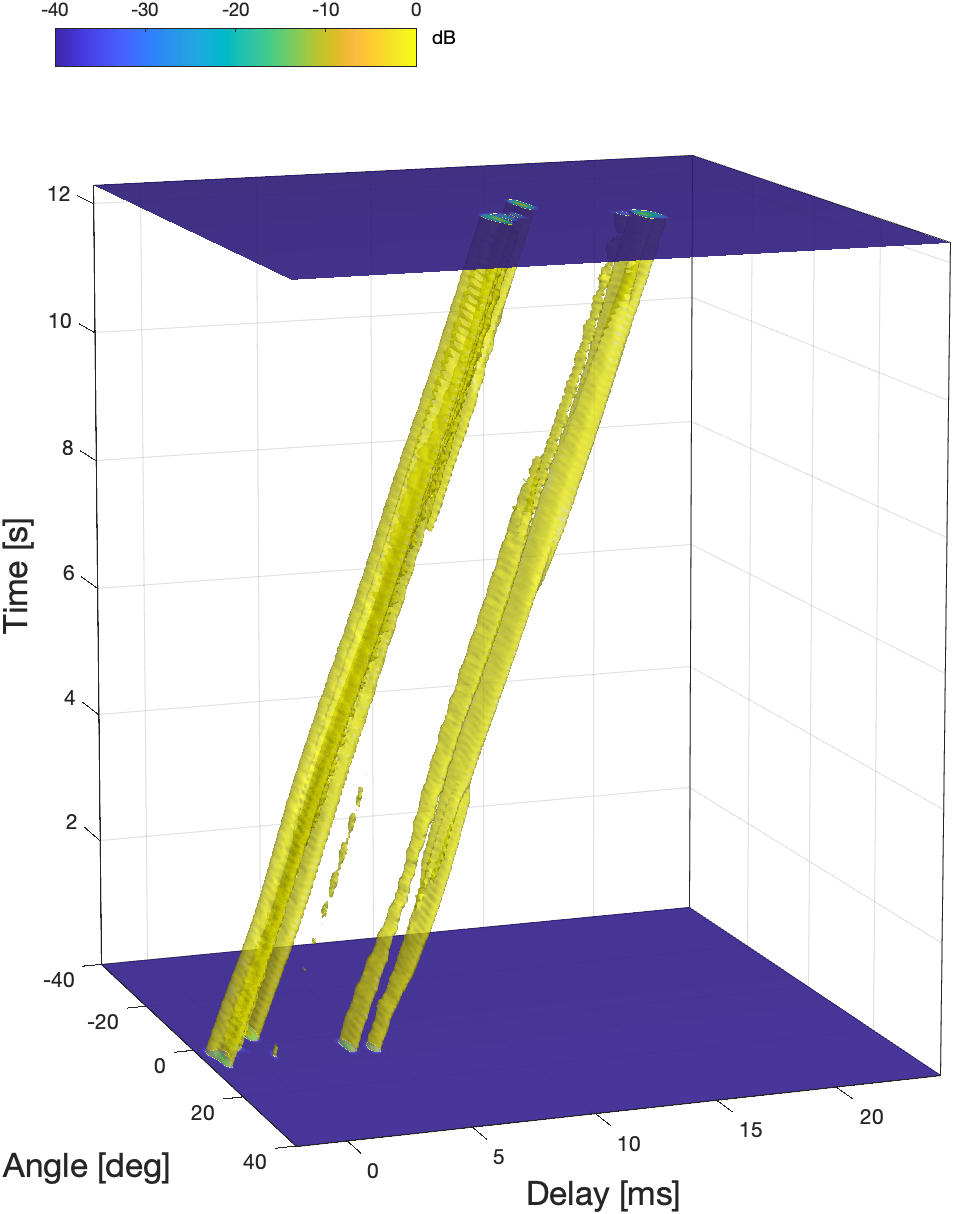}
        \caption{}
        \label{fig:delay_angle_time}
    \end{subfigure}
    \end{subfigure}
    \caption{(a) Shallow-water channel geometry illustrating the direct, surface-reflected, and bottom-reflected propagation paths between the transmitter and receiver array. (b) Three-dimensional visualization of the propagation paths observed at the receiver as functions of angle of arrival, delay, and time. The multiple yellow trajectories illustrate the multipath structure of the channel and their slanted evolution reveals time-varying delays caused by platform motion.}
    \label{fig:underwater_acoustic_channel}
\end{figure}

To illustrate these underwater channel effects, consider the shallow-water channel geometry shown in Fig.~\ref{fig:geometry}. The transmitter is initially located at a distance $L_{TR}$ from the receiver array, while the transmitter and receiver are positioned at depths $D_T$ and $D_R$, respectively. We consider three representative propagation paths: the direct path, the surface-reflected path, and the bottom-reflected path. The propagation paths arrive at the receiver with relative delays $\tau_p,\ p=0, 1, 2,$ and angles of arrival $\theta_p$ (for example, the angle of arrival of the direct path is denoted by $\theta_0$ and measured with respect to the horizontal axis). As the transmitter moves relative to the receiver with velocity $v_{tr}$, the propagation paths continuously vary their lengths, resulting in time-varying delays modeled as $\tau_p(t)=\tau_p+a_p \cdot t$, where $a_p = v_{tr,p} / c$ is the Doppler scaling factor. This scaling factor is related to the relative transmitter/receiver velocity projected onto the $p$-th path $v_{tr,p}$ and the speed of sound $c$ (nominally $1500$~m/s in water). Figure~\ref{fig:delay_angle_time} illustrates a three-dimensional visualization of the underwater acoustic propagation paths, extracted from a real recording, as a function of angle of arrival, delay, and time. The multiple yellow trajectories highlight the multipath structure of the channel, while their slanted evolution over time reveals time-varying delays caused by platform motion.


To mitigate the time-varying effects and multipath propagation, underwater acoustic receivers commonly employ array processing techniques together with adaptive equalization methods. In particular, beamforming techniques are widely used to provide spatial gain, interference suppression, path separation, and improved signal detection. When multiple strong arrivals exist, and their relative gains and delays vary across the duration of transmission, beamforming approaches often rely on estimating one or more dominant angles of arrival for stable paths and subsequently constructing beamforming filters matched to those estimated directions. Such approaches often require hard beam selection or switching operations and may become sensitive to directional estimation errors in dynamic underwater acoustic channels. Consequently, beamformers designed for static or slowly varying geometries may experience degraded performance when the spatial structure of the channel varies rapidly over time.

In this work, we propose a Bayesian adaptive beamforming framework based on a universal mixture over a finite set of steering hypotheses. The receiver maintains a set of candidate steering directions with prior probabilities representing spatial uncertainty. Each beamformer output is assigned a posterior probability that quantifies its consistency with the received observations. These probabilities are recursively updated via Bayes' theorem using exponential-family likelihoods, yielding sequential exponential-weighting updates. The beamformer outputs are then probabilistically combined, enabling joint angle-of-arrival estimation, adaptive beamforming, and continuous spatial tracking without hard beam switching.


Building upon this probabilistic framework, we further introduce a universal equalization approach in which an adaptive equalizer is placed after each beamforming branch. While the beamformers continuously track the evolving angles of arrival and spatial structure of the channel, the equalizers compensate for channel gain fluctuations, intersymbol interference, and path-dependent Doppler drifts. The posterior probabilities are similarly updated using exponential likelihood functions constructed from the branch-specific data-detection performance. The branch outputs are then combined according to their posterior probabilities, enabling joint spatio-temporal adaptation under time-varying underwater acoustic propagation conditions.

The remainder of this paper is organized as follows. Section~\ref{sec:universal_bf} introduces the proposed universal adaptive beamforming framework and derives the associated Bayesian posterior adaptation rule. A Gaussian observation model is then used to establish the connection between the probabilistic formulation and classical \ac{mmse} beamforming. Section~\ref{sec:universal_bf_equ} extends the framework toward universal adaptive equalization. Finally, experimental results and conclusions are presented in Section~\ref{sec:experiment} and Section~\ref{sec:conclusion}, respectively.
\section{Universal Adaptive Beamforming}
\label{sec:universal_bf}

We first consider a narrowband array model to introduce the proposed Bayesian beamforming framework. Let $v_m[n]$ denote the complex baseband sample observed on the $m$-th array element at time index $n$. Under the plane-wave and narrowband assumptions, the received signal can be expressed as
\begin{equation}
v_m[n] =  x[n] e^{-jm\chi_0} + w_m[n], \quad m =0,\ldots,M-1
\label{eq:bf_model}
\end{equation}
where $x[n]$ denotes the complex envelope of the signal of interest, $w_m[n]$ represents additive complex noise, and $\chi_0 = 2\pi f \frac{\delta}{c}\sin\theta_0$ denotes the spatial phase shift induced by plane-wave propagation across adjacent array elements. Here, $f$ is the signal frequency, $\delta$ is the inter-element spacing, and $\theta_0$ is the angle of arrival of the signal.


The received signal model can be equivalently expressed as
\begin{equation}
\mathbf{v}[n] = \mathbf{s}_M(\chi_0)x[n] + \mathbf{w}[n]
\label{eq:bf_vector_model}
\end{equation}
where $\mathbf{v}[n] = \begin{bmatrix}
v_0[n] &
v_1[n] &
\ldots &
v_{M-1}[n]
\end{bmatrix}^{\top}$ and 

\begin{equation}
\mathbf{s}_M(\chi) = \begin{bmatrix} 1 & e^{-j\chi} & \ldots & e^{-j(M-1)\chi}
\end{bmatrix}^{\top}
\end{equation}
denotes the steering vector associated with spatial phase shift $\chi$.

In the array signal processing literature, the tasks are twofold: to estimate the angle of arrival $\hat{\theta}_0$
and subsequently construct a beamforming filter
$\boldsymbol{\Phi}(\hat{\theta}_0)$. The resulting beamformed output provides an estimate of the desired signal
\begin{equation}
    \hat{x}[n] = \boldsymbol{\Phi}^H(\hat{\theta}_0) \mathbf{v}[n]
\end{equation}


In the proposed universal beamforming framework, angle inference and spatial filtering are performed jointly through adaptive probabilistic weighting over multiple steering hypotheses.

\subsection{Bayesian approach}

Consider a grid of hypothesized angles of arrival $\theta_i \in \Theta = \{ \theta_1, \theta_2,\ldots,\theta_K \}$. For each hypothesis, we obtain a corresponding beamforming filter $\boldsymbol{\Phi}_i = \boldsymbol{\Phi}(\theta_i)$ and compute the output
\begin{equation}
    \hat{x}_i[n] = \boldsymbol{\Phi}_i^H \mathbf{v}[n]
\end{equation}
The final estimate of the signal is expressed through a weighted blending operation
\begin{equation}
    \hat{x}[n] = \sum_{i=1}^K \mu_i[n] \hat{x}_i[n]
\end{equation}
where the weights $\mu_i[n] \triangleq P\left(\theta_i | \mathbf{v}_0^{n-1}\right)$ are interpreted as the posterior probabilities of the hypothesis angles $\theta_i$ given the observation signal samples ${\mathbf v}_0^{n-1} = \left\lbrace \mathbf{v}[0], \mathbf{v}[1],\ldots,\mathbf{v}[n-1] \right\rbrace$. These posterior probabilities are computed using Bayes' theorem
\begin{equation}
    P\left(\theta_i | \mathbf{v}_0^{n-1}\right) = \frac{P_i \cdot  P\left(\mathbf{v}_0^{n-1} | \theta_i \right)}{\sum_{j=1}^K P_j \cdot P\left(\mathbf{v}_0^{n-1} | \theta_j\right)}
\end{equation}
where $P_i$ are the prior probabilities associated with the hypothesis $\theta_i$, i.e. $P_i = P(\theta_i)$, and the conditional likelihood $P\left(\mathbf{v}_0^{n-1} | \theta_i\right)$ admits an exponential form \cite{singer1999universal,buck2018performancedmr}

\begin{equation}
    P\left(\mathbf{v}_0^{n-1} | \theta_i\right) = B \cdot  \exp \left( -\frac{1}{2\nu} \mathcal{L}\left(\mathbf{v}_0^{n-1},\boldsymbol{\Phi}_i\right) \right)
\end{equation}
where $B$ is a scaling constant, $\nu$ is a constant that
controls the sensitivity to the loss function $\mathcal{L}\left(\mathbf{v}_0^{n-1}, \boldsymbol{\Phi}_i\right)$, which depends on the input samples $\mathbf{v}_0^{n-1}$ and hypothesis $\theta_i$.

\subsubsection{Example}

Consider the scenario described by Eq. \eqref{eq:bf_vector_model} where $x[n]$ is an i.i.d. source of interest arriving from an unknown direction $\theta_0$, e.g. $x[n] \sim \mathcal{CN}(0,\sigma_x^2)$, and the noise is zero-mean and i.i.d. circularly symmetric complex Gaussian, e.g. $\mathbf{w}[n] \sim \mathcal{CN}(\mathbf{0},\mathbf{R}_w)$ with covariance $\mathbf{R}_w = \sigma_w^2 \mathbf{I}_M$.

Defining the $i$-th hypothesis as
\begin{equation}
\mathbf{v}[n] = \mathbf{s}_M\left(\chi_i\right) x[n] + \mathbf{w}[n]
\label{eq:bf_narrowband_model_example}
\end{equation}
where $\chi_i = 2\pi f \tfrac{\delta}{c} \sin \theta_i$. Under the steering hypothesis $\theta_i$, the received vector $\mathbf{v}[n]$ is distributed according to $\mathbf{v}[n] | \theta_i \sim \mathcal{CN}(\mathbf{0},\mathbf{R}_i)$, where

\begin{equation}
\mathbf{R}_i = \sigma_x^2 \mathbf{s}_M(\chi_i)\mathbf{s}_M^H(\chi_i)+\sigma_w^2 \mathbf{I}_M
\end{equation}
Then, we obtain the joint conditional probability density function upon observing $\mathbf{v}_0^{n-1}$
\begin{equation}
\begin{aligned}
P\left(\mathbf{v}_0^{n-1} | \theta_i\right) &=
\prod_{k=0}^{n-1}
p(\mathbf{v}[k]|\theta_i)=
\prod_{k=0}^{n-1}
\frac{
1
}{
\pi^M \det(\mathbf{R}_i)
}
\exp\left(
-\mathbf{v}^H[k]
\mathbf{R}_i^{-1}
\mathbf{v}[k]
\right)
\\
&=
\frac{
1
}{
\left(
\pi^M\det(\mathbf{R}_i)
\right)^n
}
\exp\left(
-\sum_{k=0}^{n-1}
\mathbf{v}^H[k]
\mathbf{R}_i^{-1}
\mathbf{v}[k]
\right)
\end{aligned}
\end{equation}
Applying the matrix inversion lemma to $\mathbf{R}_i^{-1}$ and removing terms independent of the steering hypothesis $\theta_i$, the conditional likelihood can be expressed as
\begin{equation}
P\left(\mathbf{v}_0^{n-1}|\theta_i\right)
\propto
\exp\left(
-\frac{1}{2\nu}
\mathcal{L}\left(\mathbf{v}_0^{n-1},\boldsymbol{\Phi}_i\right)
\right)
\end{equation}
with $\nu = \sigma_x^2 / \left( 2+ \tfrac{2\sigma_x^2}{\sigma_w^2} M\right)$ and a loss function defined as
        

\begin{equation}
    \mathcal{L}\left(\mathbf{v}_0^{n-1},\boldsymbol{\Phi}_i\right) = -\sum_{k=0}^{n-1} |\boldsymbol{\Phi}_i^H \mathbf{v}[k]|^2
\end{equation}
where the beamformer
\begin{equation}
    \boldsymbol{\Phi}_i = \frac{\sigma_x^2}{\sigma_w^2 + \sigma_x^2 M} \mathbf{s}_M(\chi_i)
\end{equation}
coincides with the linear \ac{mmse} beamformer associated with the Gaussian observation model\cite{bell2000bayesianbf,vantrees2002array,lam2006bayesianbf}, i.e. $\boldsymbol{\Phi}_{\text{MMSE}} = \sigma_x^2 \mathbf{R}_i^{-1} \mathbf{s}_M(\chi_i)$.

\begin{figure}[h]
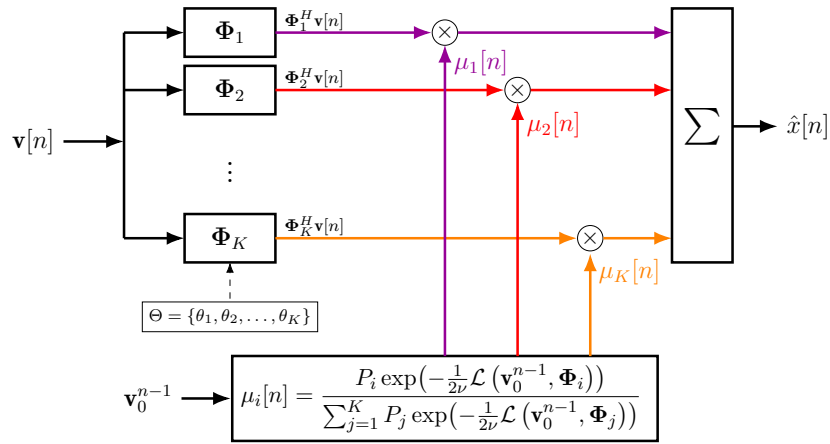

\centering
\begin{subfigure}[b]{0.8\textwidth}
    \centering
    \scalebox{0.8}{\universalbeamforming}
\end{subfigure}
\caption{Universal Adaptive Beamforming.}
\label{fig:universal_bf_diagram}
\end{figure}

The posterior weights are then computed as

\begin{equation}
    \mu_i[n] = \frac{P_i \cdot  \exp\left( - \frac{1}{2\nu} \mathcal{L}\left(\mathbf{v}_0^{n-1},\boldsymbol{\Phi}_i\right) \right)}{\sum_{j=1}^K P_j \cdot \exp\left( - \frac{1}{2\nu} \mathcal{L}\left(\mathbf{v}_0^{n-1},\boldsymbol{\Phi}_j\right) \right)}
    \label{eq:mu_weights_bf}
\end{equation}
where the denominator ensures that the weights add up to one
\begin{equation}
    \sum_{i=1}^K \mu_i[n] = 1
\end{equation}
Thus, steering hypotheses producing larger coherent beamformer output energy receive larger posterior probability and therefore contribute more strongly to the final beamformed output. In Fig.~\ref{fig:universal_bf_diagram}, we illustrate the block diagram of the universal adaptive beamforming algorithm.

A simulation example is presented to illustrate the posterior adaptation behavior of the proposed universal beamforming framework under unknown angle-of-arrival conditions. We consider the narrowband model of Eq.~\eqref{eq:bf_vector_model} with a vertical linear array of $12$ elements equally-spaced by $\delta = 0.075$~m (corresponding to half of the wavelength), the speed of sound is $c = 1500$~m/s, and the frequency of interest is $f = 10$~kHz. The signal source $x[n]$ and noise elements $w_m[n]$ are i.i.d. and drawn from zero-mean circularly-symmetric complex Gaussian distributions with variances $\sigma_x^2$ and $\sigma_w^2$. The \ac{snr} in decibels is defined as $\mathrm{SNR} = 10 \log_{10} \frac{\sigma_x^2}{\sigma_w^2}$ and set to $0$~dB. Two scenarios are considered: 1) a fixed-angle source with $\theta_0 = 11^\circ$ and 2) a time-varying source whose angle of arrival evolves over time $\theta_0[n] = 11^\circ - \tfrac{1}{50000} n$.

\begin{figure}[h]
\centering
\begin{subfigure}[b]{1\textwidth}
\begin{subfigure}[b]{0.24\textwidth}
    \centering
    \includegraphics[width=\linewidth]{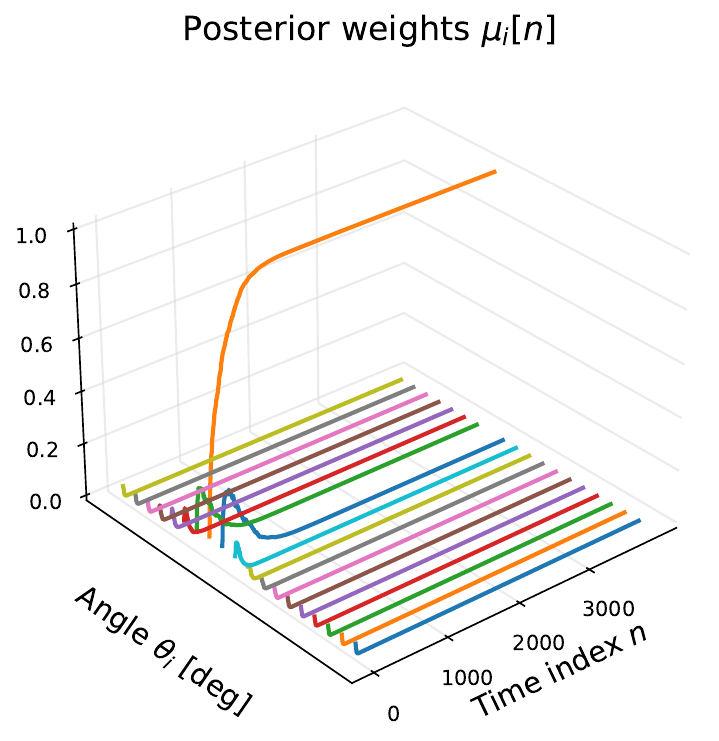}
    \caption{}
    \label{fig:posterior_weights_scenario_1}
\end{subfigure}
\begin{subfigure}[b]{0.24\textwidth}
    \centering
    \includegraphics[width=\linewidth]{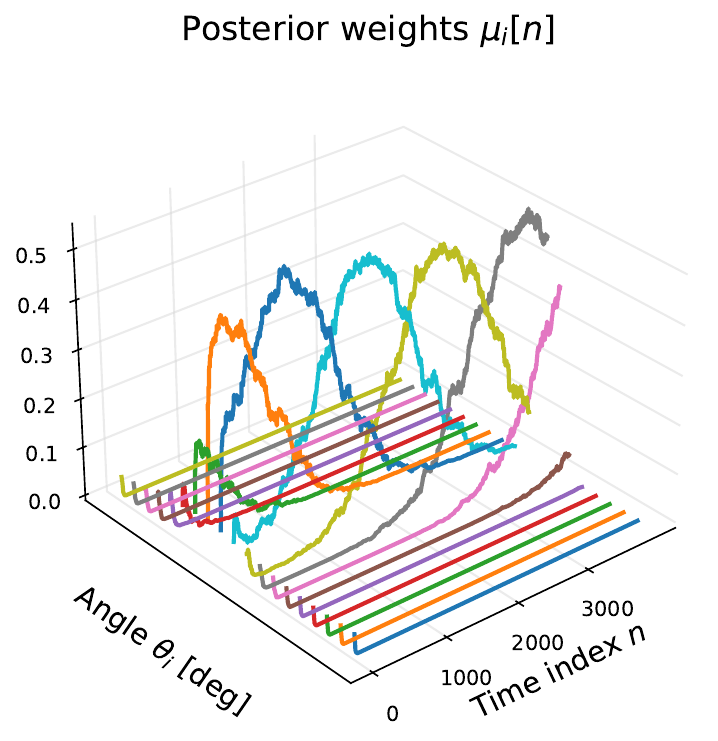}
    \caption{}
    \label{fig:posterior_weights_scenario_2}
\end{subfigure}
\begin{subfigure}[b]{0.24\textwidth}
    \centering
    \includegraphics[width=\linewidth]{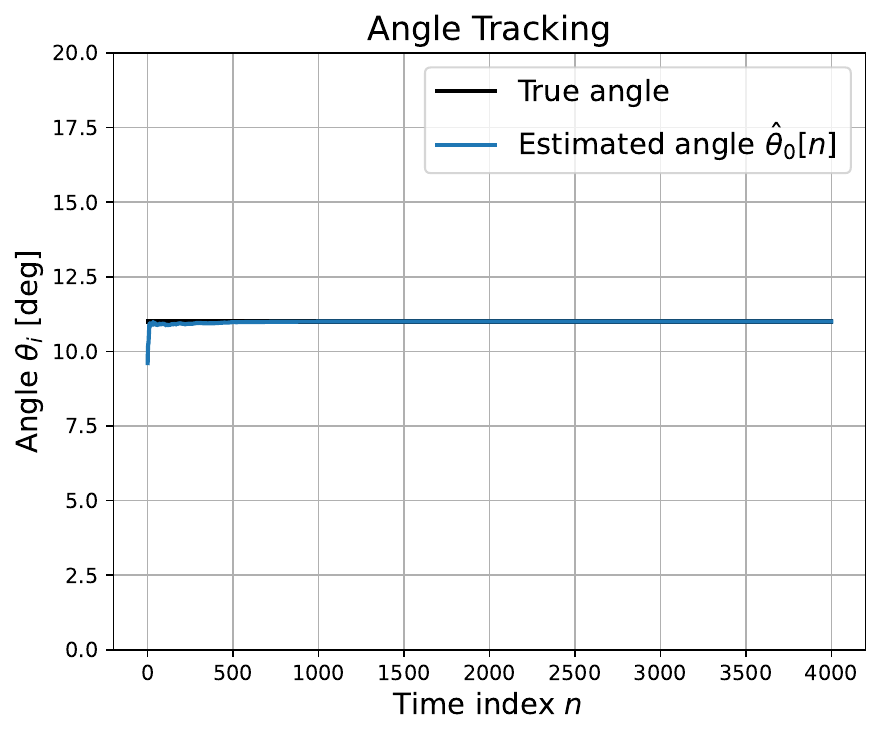}
    \caption{}
    \label{fig:angle_estimation_1}
\end{subfigure}
\begin{subfigure}[b]{0.24\textwidth}
    \centering
    \includegraphics[width=\linewidth]{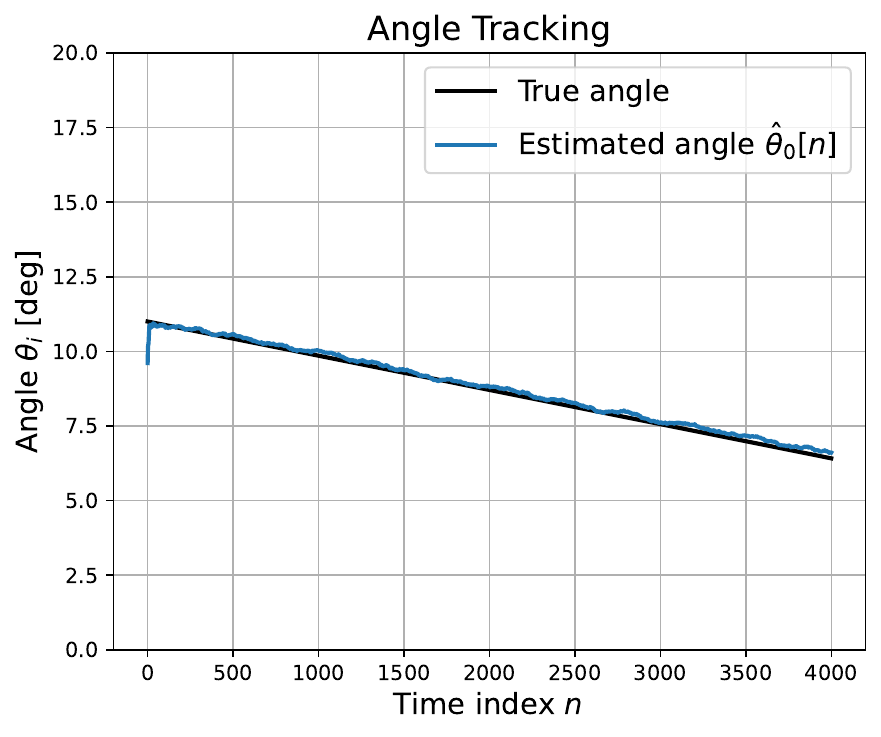}
    \caption{}
    \label{fig:angle_estimation_2}
\end{subfigure}
\end{subfigure}
\caption{(a,b) Evolution of the posterior weights $\mu_i[n]$. (c,d) Angle of arrival estimation $\hat{\theta}_0[n] = \sum_{i=1}^K \mu_i[n] \cdot \theta_i$.}
\label{fig:sim_posterior_weights}
\end{figure}

The hypothesis angles are set to $\Theta \in [0^\circ,19^\circ]$ and uniformly distributed according to $P_i = 1/20$, $\forall i$. In Fig.~\ref{fig:posterior_weights_scenario_1} and Fig.~\ref{fig:posterior_weights_scenario_2}, we illustrate the evolution of the posterior weights $\mu_i[n]$ for both cases. For the static scenario, the posterior distribution rapidly converges toward the correct steering direction $\theta_0$, producing a stable angle estimate and coherent beamformer output. In the time-varying scenario, the posterior weights continuously redistribute across neighboring steering hypotheses as the source direction changes, demonstrating the ability of the proposed framework to perform soft spatial tracking without explicit beam switching.

Simultaneous tracking of the angle of arrival can be established by computing
\begin{equation}
    \hat{\theta}_0[n] = \sum_{i=1}^K \mu_i[n] \cdot  \theta_i
\end{equation}
and the performance is shown in Fig.~\ref{fig:angle_estimation_1} and Fig.~\ref{fig:angle_estimation_2} for both scenarios. We observe that the resulting angle estimates closely follow the true angle $\theta_0[n]$, illustrating the adaptive nature of the Bayesian posterior weighting mechanism.


\section{Universal Beamforming and Equalization}
\label{sec:universal_bf_equ}

Following the idea of universal beamforming and prediction, we extend the application to underwater acoustic communication and equalization. In this case, the baseband continuous-time received signal on the $m$-th element is given by
\begin{equation}
    v_m(t) = \sum_n d[n] \, \underline{h}_m(t,t-nT) + w_m(t)
\end{equation}
where $w_m(t)$ is additive noise, $d[n]$ represents the data symbols taken from a constellation, e.g. PSK, and $\underline{h}_m(t,\tau)$ represents the baseband-equivalent time-varying channel response. The variable $t$ denotes the observation time associated with the temporal evolution of the propagation environment, while $\tau$ denotes the delay variable of the impulse response. The baseband-equivalent channel response is modeled as
\begin{equation}
    \underline{h}_m(t,\tau) = \sum_{p=0}^{P-1} h_p^m(t) g(\tau- \tau_p^m(t)) e^{-j2\pi f_c \tau_p^m(t)}
\end{equation}
where $h_p^m(t)$, $\tau_p^m(t)$, and $P$ denote the path gains, path delays, and total number of propagation paths, respectively. The carrier frequency is represented by $f_c$, while $g(t)$ denotes the transmitted pulse-shaping filter, e.g. a raised-cosine pulse with roll-off factor $\alpha_{rc}$. The symbol duration is denoted by $T = 1/R$, which is associated with the symbol rate $R$. The baseband-equivalent impulse response may be interpreted as a pulse-shaped version of the physical time-varying impulse response
\begin{equation}
    h_m(t,\tau) = \sum_{p=0}^{P-1} h_p^m(t) \delta(\tau- \tau_p^m(t))
\end{equation}

The time-varying propagation delays are modeled according to
\begin{equation}
\tau_p^m(t) = \tau_p^m + \epsilon_p(t), \quad m=0,\ldots,M-1
\end{equation}
where $\epsilon_p(t)$ represents the Doppler-induced perturbation associated with the $p$-th propagation path. For sufficiently small array apertures, the Doppler perturbation is assumed approximately identical across sensors for each path.

Assuming plane-wave propagation, the propagation delays across the array aperture satisfy
\begin{equation}
\tau_p^m = \tau_p^0 + m\frac{\delta}{c}\sin\theta_p, \quad m=0,\ldots,M-1
\end{equation}
where $\theta_p$ is the angle of arrival associated with the $p$-th propagation path. The proposed receiver employs a front-end broadband beamformer to spatially align the received array observations across the occupied signal bandwidth prior to equalization. The beamformed branch signals are subsequently processed by adaptive equalizers that jointly compensate residual Doppler distortion, intersymbol interference, and time-varying channel fluctuations in order to detect the transmitted data symbols.




\subsection{Broadband Beamforming}

The narrowband beamforming model introduced in Section~\ref{sec:universal_bf} assumes that the signal bandwidth is sufficiently small such that spatial propagation delays across the array can be represented by frequency-independent phase shifts. In underwater acoustic communication systems, however, the transmitted waveform occupies a finite bandwidth and therefore requires frequency-dependent spatial filtering. Because the propagation delay across the array depends on frequency, broadband spatial filtering cannot generally be implemented using a single narrowband steering vector. Instead, frequency-dependent beamforming responses must be synthesized across the occupied signal bandwidth.

For each steering hypothesis $\theta_i$, we define a broadband beamforming response
\begin{equation}
\boldsymbol{\Phi}_i(f)
=
\begin{bmatrix}
\phi_{i,0}(f) &
\phi_{i,1}(f) &
\ldots &
\phi_{i,M-1}(f)
\end{bmatrix}^{\top}
\label{eq:bf_broadband_vector}
\end{equation}
where the explicit dependence on frequency is emphasized in the response associated with the $m$-th array element
\begin{equation}
\phi_{i,m}(f) = \exp\left(-j2\pi f m\tfrac{\delta}{c}\sin\theta_i \right)
\end{equation}

The beamforming filters are synthesized over a discrete set of frequency bins spanning the occupied signal bandwidth. Let the frequency bins be defined as
\begin{equation}
    f_\ell =
    \begin{cases}
    f_c + \ell \Delta f,
    & 0 \le \ell \le \frac{L}{2}
    \\[2mm]
    f_c + (\ell-L)\Delta f,
    & \frac{L}{2}+1 \le \ell \le L-1
    \end{cases}
\end{equation}
where $L$ is the total number of frequency bins and $\Delta f = \tfrac{1}{L T_s}$ is the frequency resolution associated with the sampling interval $T_s$. For each steering hypothesis $\theta_i$, the desired broadband spatial response is sampled over the discrete frequency grid according to
\begin{equation}
\boldsymbol{\Phi}_{i,\ell} = \boldsymbol{\Phi}_i(f_\ell) = \begin{bmatrix} \phi_{i,0}[\ell] & \phi_{i,1}[\ell] & \ldots & \phi_{i,M-1}[\ell] \end{bmatrix}^{\top}
\end{equation}
where $\phi_{i,m}[\ell] = \phi_{i,m}(f_\ell)$. In addition, the synthesized beamforming responses are set to zero for the frequencies outside of the signal band
\begin{equation}
    \boldsymbol{\Phi}_{i,\ell} =
    \begin{cases}
    \boldsymbol{\Phi}_i(f_\ell),
    & \ell\in\mathcal{L}_B \\[2mm]
    \mathbf{0}, & \ell\notin\mathcal{L}_B
    \end{cases}
\end{equation}
where the subset of active frequency bins is defined as
\begin{equation}
\mathcal{L}_B = \left\{
\ell \;\middle|\; 0 \le \ell \le \bar{L}
\;\; \text{or} \;\; L-\bar{L} \le \ell \le L-1 \right\}
\end{equation}
with $\bar{L}
=
\left\lceil
\tfrac{L(1+\alpha_{rc})}{2N_s}
\right\rceil$, where $N_s=T/T_s$ is the oversampling ratio.

The broadband beamforming frequency responses are subsequently transformed into realizable time-domain FIR filters through an \ac{idft}. Before applying the \ac{idft}, we obtain $\tilde{\phi}_{i,m}[\ell] = (-1)^\ell \phi_{i,m}^*[\ell]$,
where the factor $(-1)^\ell$ introduces an $L/2$-sample circular shift of the resulting impulse response, thereby centering the synthesized FIR filter. This operation is equivalent to applying the same shift directly in the time domain after the \ac{idft}. Specifically, the synthesized beamforming filter for the $m$-th array element and the $i$-th steering hypothesis is given by
\begin{equation}
\psi_{i,m}(nT_s) =  \frac{1}{L} \sum_{\ell=0}^{L-1}
\tilde{\phi}_{i,m}[\ell] e^{j2\pi \ell n/L} = \frac{1}{L}
\sum_{\ell\in \mathcal{L}_B} \tilde{\phi}_{i,m}[\ell] e^{j2\pi \ell n/L}.
\end{equation}

The synthesized filters compensate for frequency-dependent propagation delays across the array aperture and therefore implement broadband spatial alignment over the signal bandwidth. The broadband beamformed signal associated with steering hypothesis $\theta_i$ is therefore expressed as
\begin{equation}
\tilde{y}_i(nT_s)
=
\sum_{m=0}^{M-1}
\psi_{i,m}(nT_s) \star
\,v_m(nT_s)
\label{eq:y_i_bf}
\end{equation}
where $\star$ represents convolution. The resulting branch signals $\tilde{y}_i(nT_s)$, $i=1,\ldots,K$, are subsequently processed by adaptive equalizers to compensate for Doppler-induced effects and to detect the transmitted data symbols.

\subsection{Universal Equalization}

For each steering hypothesis $\theta_i$, the beamformed branch signal $\tilde{y}_i[n]$ is processed by a fractionally-spaced \ac{dfe} operating at spacing $T_s=T/2$. The equalizer jointly compensates residual Doppler distortion, intersymbol interference, and time-varying channel fluctuations in order to recover the transmitted data symbols.

The Doppler compensation is performed through two coupled operations: (1) adaptive resampling to compensate for the time-varying delays, and (2) residual phase tracking using a digital \ac{pll}.

Adaptive resampling is applied to the signal in Eq.~\eqref{eq:y_i_bf} using linear interpolation. Specifically, during the $n$-th symbol interval, two interpolated samples are generated according to
\begin{equation}
    \begin{aligned}
        y_i\left(nT+\ell^\prime\frac{T}{2}\right) &= \mathcal{I} \left\lbrace \tilde{y}_i\left(nT + \ell^\prime\frac{T}{2} - \frac{\hat{\varphi}_i(nT)}{2\pi f_c}\right)\right\rbrace, \quad \ell^\prime=N_1, N_1 -1
    \end{aligned}
\end{equation}
where $\mathcal{I}\{\cdot\}$ denotes linear interpolation as described in \cite{cuji2026path}, $N_1$ represents the number of causal feedforward taps, and $\hat{\varphi}_i(nT)$ denotes the estimated Doppler-induced carrier phase associated with the $i$-th branch.

The interpolated samples are inserted into the top of an existing vector $\mathbf{y}_i[n-1]$ and the two bottom elements of $\mathbf{y}_i[n-1]$ are discarded (represented by $\mathbf{y}_i[n-1]_{1:N_f-2}$). The resulting equalizer input vector of length $N_f$ is therefore given by
\begin{equation}
\mathbf{y}_i[n]
=
\begin{bmatrix}
y_i\left(nT+N_1\frac{T}{2}\right)
\\
y_i\left(nT+N_1\frac{T}{2}-\frac{T}{2}\right)
\\
\mathbf{y}_i[n-1]_{1:N_f-2}
\end{bmatrix},
\qquad
i=1,\ldots,K
\label{eq:y_p}
\end{equation}

The \ac{dfe} structure follows the formulation proposed in \cite{stojanovic1993multichanneldfe}. A feedforward filter $\mathbf{a}_i[n]$, consisting of $N_f$ coefficients, is applied to the input vector $\mathbf{y}_i[n]$ and phase-corrected according to the estimated carrier phase
\begin{equation}
x_i[n] = \mathbf{a}_i^H[n] \mathbf{y}_i[n] e^{-j\hat{\varphi}_i(nT)}.
\end{equation}


The feedback equalizer filter $\mathbf{b}_i[n]$, which contains $N_b$ coefficients, operates on the vector of symbol decisions 
$$\tilde{\mathbf{d}}_i[n] = \begin{bmatrix}
    \tilde{d}_i[n-1] & \tilde{d}_i[n-2] & \ldots & \tilde{d}_i[n-N_b]
\end{bmatrix}^\top$$
to produce $z_i[n] = \mathbf{b}_i^H [n] \tilde{\mathbf{d}}_i[n]$. The decision $\tilde{d}_i[n] = \mathrm{decision} \left\lbrace\hat{d}_i[n]\right\rbrace$ is formed by quantizing the estimate $\hat{d}_i[n]$ to the nearest symbol value. The estimate of the data symbol is then obtained as
\begin{equation}
    \hat{d}_i[n] = x_i[n] - z_i[n] = \mathbf{c}_i^H[n] \mathbf{u}_i[n], \quad i=1,\ldots,K
\end{equation}
where the composite vector of the feedforward and feedback equalizer filter taps is formed as
\begin{equation}
    \mathbf{c}_i[n] = \begin{bmatrix}
    \mathbf{a}_i^\top[n] & -\mathbf{b}_i^\top[n]
\end{bmatrix}^\top
\end{equation}
and the composite vector of signal samples is given by
\begin{equation}
    \begin{aligned}
    \mathbf{u}_i[n] &= \big[
    \mathbf{y}_i^\top[n] e^{-j \hat{\varphi}_i(nT)} \;\;  \tilde{\mathbf{d}}_i^\top[n] \big]^\top
    \end{aligned}
\end{equation}
and the phases $\hat{\varphi}_i(nT)$ are updated using a digital \ac{pll}
\begin{equation}
    \begin{aligned}
        \hat{\varphi}_i(nT + T) = &\hat{\varphi}_i(nT) + K_{f_1} \xi_i[n] + K_{f_2} \sum_{q=0}^n \xi_i[q], \quad i=1,\ldots,K
    \end{aligned}
\end{equation}
where $\xi_i[n] = \mathrm{Im} \left\lbrace x_i[n] [\tilde{d}_i[n] + z_i[n]]^* \right\rbrace$, $K_{f_1}$ and $K_{f_2}$ represent the \ac{pll} tracking coefficients, and the phase estimates are initialized as $\hat{\varphi}_i(0)=0$, $i=1,\ldots,K$.
The symbol error 
\begin{equation}
e_i[n] =
\begin{cases}
d[n] - \hat{d}_i[n], & \text{training mode}, \\[4pt]
\tilde{d}_i[n] - \hat{d}_i[n], & \text{decision-directed mode},
\end{cases}
\end{equation}
is then used to obtain the weights $\mathbf{c}_i[n]$ recursively over time as
\begin{equation}
    \mathbf{c}_i[n+1] = \mathbf{c}_i[n] + \mathcal{A}\left\lbrace \mathbf{u}_i[n], e_i[n] \right\rbrace
    \label{eq:dfe_coeff}
\end{equation}
where $\mathcal{A}\left\lbrace \cdot , \cdot  \right\rbrace$ denotes an update term, which can be computed using an adaptive algorithm, such as \ac{lms} or \ac{rls}, that operates on the input $\mathbf{u}_i[n]$ and the error $e_i[n]$. During training mode, the vector $\tilde{\mathbf{d}}_i[n]$ contains the training data symbols and in the decision-directed mode, the vector $\tilde{\mathbf{d}}_i[n]$ contains the symbol decisions.

\begin{figure}[h]
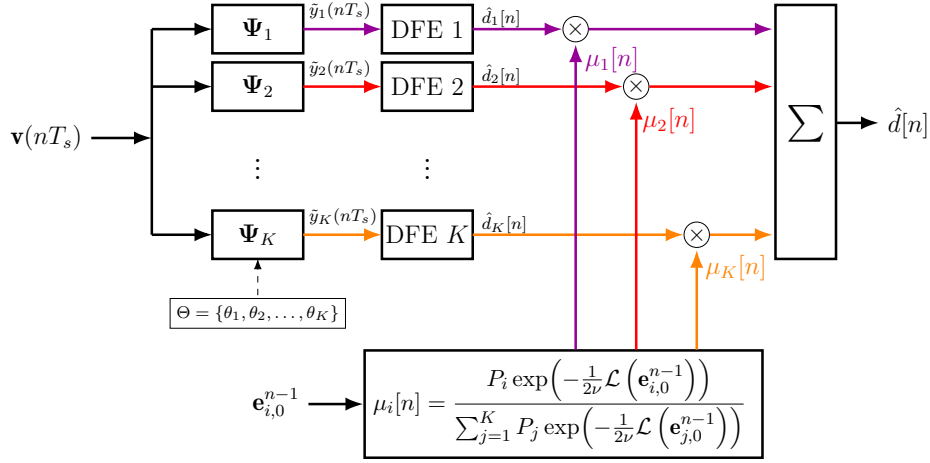

\centering
\begin{subfigure}[b]{1\textwidth}
    \centering
    \scalebox{0.8}{\universalequalizer}
\end{subfigure}
\caption{Universal Adaptive Beamforming and Equalization.}
\label{fig:universal_bfequ_diagram}
\end{figure}

The universal receiver forms the final symbol estimate by posterior-weighted blending,
\begin{equation}
    \hat{d}[n]
    =
    \sum_{i=1}^{K}
    \mu_i[n]\hat{d}_i[n].
\end{equation}
The posterior weights are similarly computed as
\begin{equation}
    \mu_i[n] = \frac{P_i \cdot  \exp\left( - \frac{1}{2\nu} \mathcal{L}\left(\mathbf{e}_{i,0}^{n-1}\right) \right)}{\sum_{j=1}^K P_j \cdot \exp\left( - \frac{1}{2\nu} \mathcal{L}\left(\mathbf{e}_{j,0}^{n-1}\right) \right)}
\end{equation}
where the loss function is now defined as
\begin{equation}
    \mathcal{L}\left(\mathbf{e}_{i,0}^{n-1}\right) = \sum_{k=0}^{n-1} |e_i[k]|^2
\end{equation}
with argument $\mathbf{e}_{i,0}^{n-1} = \lbrace e_i[0],e_i[1],\ldots,e_i[n-1] \rbrace$. Thus, branches producing smaller equalization error receive larger posterior probability and contribute more strongly to the final symbol estimate. A block diagram of the universal adaptive beamforming and equalization receiver is illustrated in Fig.~\ref{fig:universal_bfequ_diagram}.

\section{Experimental Results}
\label{sec:experiment}

In this section, we present results using a recording from the 2010 Mobile Acoustic Communications Experiment (MACE). The receiver array consisted of 12 equally spaced elements and spanned a total aperture of 1.32~m. The array was deployed at a depth of 40~m, while the transmitter was towed at depths ranging from 40 to 60~m. During the experiment, the transmitter moved both away from and toward the receiver array at speeds varying between approximately 0.5 and 1.5~m/s. Each transmission lasted approximately one minute and consisted of 128 repeated \ac{bpsk}-modulated M-sequences of length 2047. The corresponding signal parameters are summarized in Table~\ref{tab:experiment_params}.

\begin{table}
\centering
\caption{MACE Signal Parameters}
\begin{center}
\begin{tabular}{c|c}
\textbf{Parameter} & \textbf{Value} \\
\hline
center frequency $f_c$ [kHz]  & 13\\
sampling frequency $f_s$ [kHz] & $10^7/256$\\
symbol rate $R$ [symbols/s] & $4882.8$\\
M-sequence length  & $2^{11}-1$\\
modulation  & BPSK \\
number of array elements $M$ & 12\\
element spacing $\delta$ [m] & 0.12\\
\end{tabular}
\label{tab:experiment_params}
\end{center}
\end{table}

The performance is evaluated in terms of the data-detection \ac{mse}
\begin{equation}
\mathrm{MSE}
=
\frac{1}{N_d-N_t}
\sum_{n=N_t}^{N_d-1}
\left|d[n]-\hat{d}[n]\right|^2
\end{equation}
where $N_t$ denotes the number of training symbols and $N_d$ denotes the total number of symbols processed in a frame.

We process a frame composed of eight M-sequences, resulting in a total of $N_d = 8(2^{11}-1)=16376$ symbols. The receiver employs an \ac{lms}-based decision-feedback equalizer with feedforward and feedback filter lengths of $N_f=15$ and $N_b=8$, respectively. The number of training symbols is set to $ N_t = 20(N_f+N_b)=460$, which corresponds to a pilot overhead of only 2.81\%. The \ac{pll} gains are chosen as $K_{f_1}=0.01$ and $K_{f_2}=K_{f_1}/10=0.001$, while the \ac{lms} step size is set to $\mu=0.01$.

\begin{figure}[h]
\centering
\begin{subfigure}[b]{1\textwidth}
\centering
\begin{subfigure}[b]{0.32\textwidth}
    \centering
    \includegraphics[width=\linewidth]{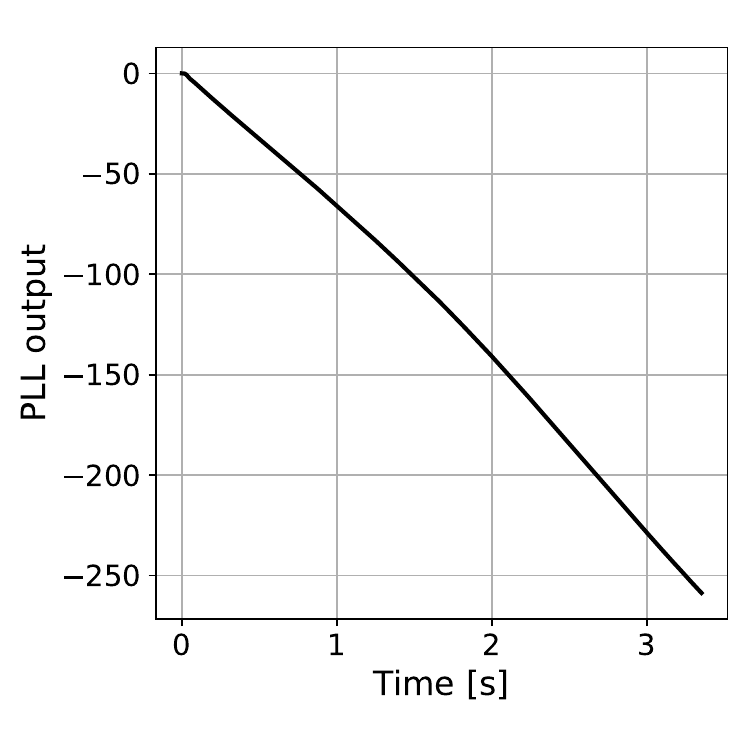}
    \caption{}
\end{subfigure}
\begin{subfigure}[b]{0.32\textwidth}
    \centering
    \includegraphics[width=\linewidth]{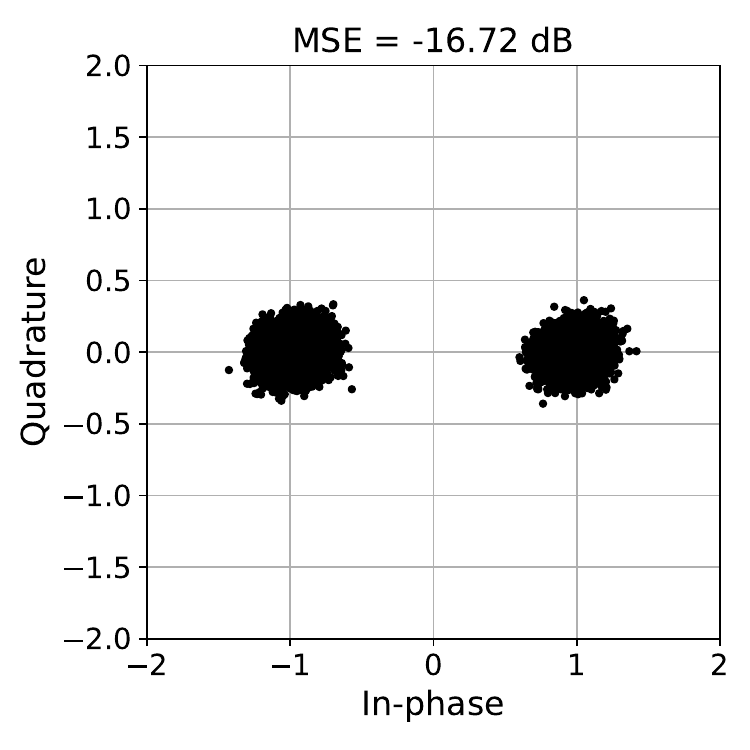}
    \caption{}
\end{subfigure}
\begin{subfigure}[b]{0.32\textwidth}
    \centering
    \includegraphics[width=\linewidth]{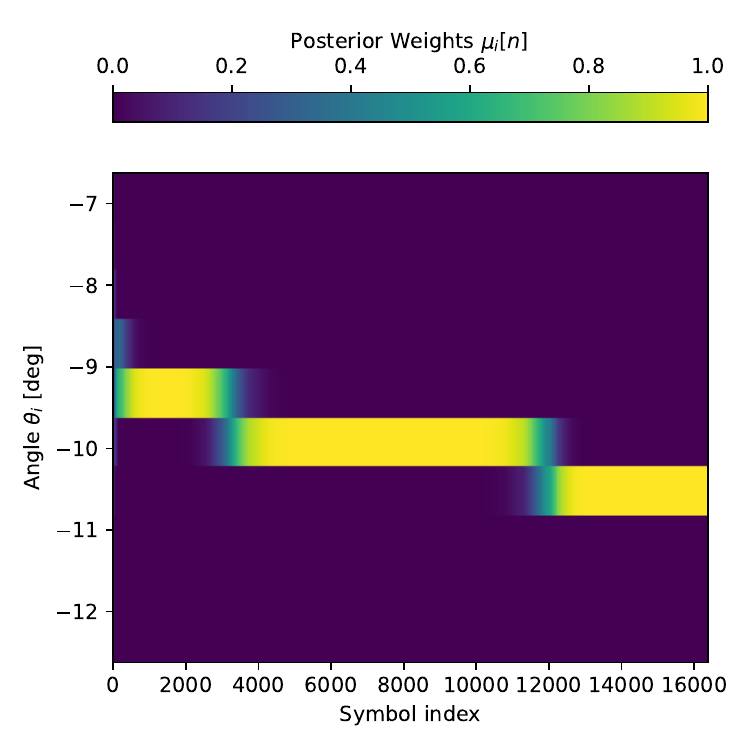}
    \caption{}
\end{subfigure}
\end{subfigure}

\caption{(a) \ac{pll} output shows the Doppler-induced phase tracking. (b) Constellation of the detected data symbols. (c) Evolution of the posterior blending weights $\mu_i[n]$.}
\label{fig:experiment}
\end{figure}

In addition, we use $K=10$ uniformly spaced steering hypotheses spanning the angular interval $ \Theta \in
[-12^\circ,-7^\circ]$, and assume a discrete uniform prior distribution over the steering hypothesis set $P_i = \frac{1}{K}$, $i=1,\ldots,K$. In Fig.~\ref{fig:experiment}, we illustrate the \ac{pll} output $\hat{\varphi}(nT) = \sum_{i=1}^K \mu_i[n] \cdot  \hat{\varphi}_i(nT)$, the constellation of the detected data symbols $\hat{d}[n]$, and the evolution of the posterior blending weights. An \ac{mse} of $-16.72$~dB and zero bit errors are reported. As observations are accumulated and branch-specific equalization errors are recursively evaluated, the posterior probabilities progressively concentrate around the steering hypotheses yielding improved communication performance. These results demonstrate the ability of the proposed receiver to continuously adapt its spatial confidence distribution according to evolving channel conditions while jointly performing adaptive beamforming and equalization under time-varying multipath propagation and Doppler distortion.

\section{Conclusion}
\label{sec:conclusion}

In this paper, we introduced a Bayesian universal adaptive beamforming framework for underwater acoustic receivers operating under dynamic propagation conditions. The proposed approach performs probabilistic model averaging across multiple steering hypotheses and continuously adapts posterior confidence according to branch-specific loss functions, enabling adaptive spatial-temporal tracking. The framework was further extended toward broadband underwater acoustic communication through frequency-domain beamformer synthesis and adaptive decision-feedback equalization. Posterior probabilities were recursively updated according to branch-specific equalization errors, allowing the receiver to jointly track angle of arrival, Doppler-induced distortions, and compensate intersymbol interference. Experimental results using MACE data demonstrated reliable communication performance together with low detection error and zero observed bit errors in a mobile scenario. The results illustrate the potential of probabilistic spatial-temporal adaptation for underwater acoustic communication in rapidly evolving propagation environments.

\bibliographystyle{IEEEtran}
\bibliography{ref}

\end{document}